\documentclass[a4paper,twocolumn,english,showpacs,aps,prl,fleqn,superscriptaddress]{revtex4-1}

\usepackage[T1]{fontenc}
\usepackage[utf8]{luainputenc}
\usepackage{amsmath}
\usepackage{amssymb}
\usepackage{graphicx}
\usepackage{hyperref}
\usepackage{babel}
\usepackage{wasysym}
\usepackage{microtype}
\usepackage{color}

\begin{document}
\renewcommand\figurename{FIG.}
\title{Strongly Interacting Bose Gases near a $d$-wave Shape Resonance}

\author{Xing-Can Yao*}
\affiliation{Shanghai Branch, National Laboratory for Physical Sciences at Microscale and Department of Modern Physics, University of Science and Technology of China, Shanghai, 201315, China}
\affiliation{CAS Center for Excellence and Synergetic Innovation Center in Quantum Information and Quantum Physics, University of Science and Technology of China, Hefei, Anhui 230026, China}
\affiliation{CAS-Alibaba Quantum Computing Laboratory, Shanghai, 201315, China}

\author{Ran Qi*}
\affiliation{Department of Physics, Renmin University of China, Beijing, 100872, China}

\author{Xiang-Pei Liu}
\affiliation{Shanghai Branch, National Laboratory for Physical Sciences at Microscale and Department of Modern Physics, University of Science and Technology of China, Shanghai, 201315, China}
\affiliation{CAS Center for Excellence and Synergetic Innovation Center in Quantum Information and Quantum Physics, University of Science and Technology of China, Hefei, Anhui 230026, China}
\affiliation{CAS-Alibaba Quantum Computing Laboratory, Shanghai, 201315, China}

\author{Xiao-Qiong Wang}
\affiliation{Shanghai Branch, National Laboratory for Physical Sciences at Microscale and Department of Modern Physics, University of Science and Technology of China, Shanghai, 201315, China}
\affiliation{CAS Center for Excellence and Synergetic Innovation Center in Quantum Information and Quantum Physics, University of Science and Technology of China, Hefei, Anhui 230026, China}
\affiliation{CAS-Alibaba Quantum Computing Laboratory, Shanghai, 201315, China}

\author{Yu-Xuan Wang}
\affiliation{Shanghai Branch, National Laboratory for Physical Sciences at Microscale and Department of Modern Physics, University of Science and Technology of China, Shanghai, 201315, China}
\affiliation{CAS Center for Excellence and Synergetic Innovation Center in Quantum Information and Quantum Physics, University of Science and Technology of China, Hefei, Anhui 230026, China}
\affiliation{CAS-Alibaba Quantum Computing Laboratory, Shanghai, 201315, China}

\author{Yu-Ping Wu}
\affiliation{Shanghai Branch, National Laboratory for Physical Sciences at Microscale and Department of Modern Physics, University of Science and Technology of China, Shanghai, 201315, China}
\affiliation{CAS Center for Excellence and Synergetic Innovation Center in Quantum Information and Quantum Physics, University of Science and Technology of China, Hefei, Anhui 230026, China}
\affiliation{CAS-Alibaba Quantum Computing Laboratory, Shanghai, 201315, China}

\author{Hao-Ze Chen}
\affiliation{Shanghai Branch, National Laboratory for Physical Sciences at Microscale and Department of Modern Physics, University of Science and Technology of China, Shanghai, 201315, China}
\affiliation{CAS Center for Excellence and Synergetic Innovation Center in Quantum Information and Quantum Physics, University of Science and Technology of China, Hefei, Anhui 230026, China}
\affiliation{CAS-Alibaba Quantum Computing Laboratory, Shanghai, 201315, China}

\author{Peng Zhang}
\affiliation{Department of Physics, Renmin University of China, Beijing, 100872, China}
\affiliation{Beijing Key Laboratory of Opto-electronic Functional Materials \& Micro-nano Devices, 100872 (Renmin Univeristy of China)}

\author{Hui Zhai}
\affiliation{Institute for Advanced Study, Tsinghua University, Beijing, 100084, China}
\affiliation{Collaborative Innovation Center of Quantum Matter, Beijing, 100084, China}

\author{Yu-Ao Chen}
\affiliation{Shanghai Branch, National Laboratory for Physical Sciences at Microscale and Department of Modern Physics, University of Science and Technology of China, Shanghai, 201315, China}
\affiliation{CAS Center for Excellence and Synergetic Innovation Center in Quantum Information and Quantum Physics, University of Science and Technology of China, Hefei, Anhui 230026, China}
\affiliation{CAS-Alibaba Quantum Computing Laboratory, Shanghai, 201315, China}

\author{Jian-Wei Pan}
\affiliation{Shanghai Branch, National Laboratory for Physical Sciences at Microscale and Department of Modern Physics, University of Science and Technology of China, Shanghai, 201315, China}
\affiliation{CAS Center for Excellence and Synergetic Innovation Center in Quantum Information and Quantum Physics, University of Science and Technology of China, Hefei, Anhui 230026, China}
\affiliation{CAS-Alibaba Quantum Computing Laboratory, Shanghai, 201315, China}


\begin{abstract}

Many unconventional quantum matters, such as fractional quantum Hall effect and $d$-wave high-Tc superconductor, are discovered in strongly interacting systems. Understanding quantum many-body systems with strong interaction and the unconventional phases therein is one of the most challenging problems in physics nowadays. Cold atom systems possess a natural way to create strong interaction by bringing the system to the vicinity of a scattering resonance. Although this has been a focused topic in cold atom physics for more than a decade, these studies have so far mostly been limited for $s$-wave resonance. Here we report the experimental observation of a broad $d$-wave shape resonance in degenerate ${}^{41}$K gas. We further measure the molecular binding energy that splits into three branches as a hallmark of $d$-wave molecules, and find that the lifetime of this many-body system is reasonably long at strongly interacting regime. From analyzing the breathing mode excited by ramping through this resonance, it suggests that a quite stable low-temperature atom and molecule mixture is produced. Putting all the evidence together, our system offers great promise to reach a $d$-wave molecular superfluid.

\end{abstract}

\maketitle

\date{\today}

In the last century, conventional quantum matters have been well understood by Landau's Fermi liquid theory, concept of symmetry breaking, as well as the perturbation theory~\cite{Landau}. Nowadays it is of great interest to find new quantum matters beyond the conventional paradigm, which can also have potential application to the next generation of quantum technology. Usually in such systems the interaction between particles is so strong that perturbation theory fails, and the system flows to a phase that is intrinsically different from a non-interacting or weakly interacting system. Ultracold atomic gases provide ideal physical systems to search for such phases, because the interaction between atoms can be precisely controlled and tuned to be very strong by bringing the system to the vicinity of a scattering resonance, for example, by the technique of magnetic Feshbach resonance ~\cite{Feshbach,Manybody}.

Indeed, in the past decades, many studies have been focused on strongly interacting atomic Bose and Fermi gases near an $s$-wave resonance, where interesting physics such as emergent scale invariance and perfect fluidity have been observed~\cite{thermodynamics0,thermodynamics1,thermodynamics2,thermodynamics3,thermodynamics4,thermodynamics5,transport0,transport1,transport2,transport3,transport4,transport5,scale2,Efimov,fluid}. On the other hand, studies so far are mostly limited to $s$-wave resonances. Only recently few works have experimentally studied many-body properties of interacting fermions near a $p$-wave resonance~\cite{pwave1}. Strongly interacting atomic gases have so far not been studied for a $d$-wave resonance and not for bosons with any high partial wave resonances. The challenge is due to the short lifetime and the narrow resonance width for higher partial waves resonances. In fact, all $d$-wave resonances reported before are quite narrow~\cite{dwave,dwave2}.

A scattering resonance can also be distinguished into a Feshbach resonance and a shape resonance, based on the origin of the bound state ~\cite{Feshbach}. For a Feshbach resonance, the bound state comes from a different channel, such as atoms in different hyperfine spin states, and it is tuned to the scattering threshold by magnetic field due to the difference in magnetic moment between this bound state and scattering states. While for a shape resonance, the bound state is supported by the attractive potential of the same channel as the scattering atoms, which in this case is the van de Waals potential between two atoms.

Magnetic field sensitivity of the bound state in a Feshbach resonance is a double-edged sword. On the one hand, most scattering resonances in cold atom system are Feshbach resonances because of its magnetic field tunability; on the other hand, once the coupling channels are weak such as in a high partial wave resonance, the resonance becomes quite narrow in terms of magnetic field, and consequently it becomes hard to investigate experimentally. In contrast, the shape resonance only occurs incidentally, but once it happens, the bound state is much less sensitive to the magnetic field and therefore it appears quite broad in magnetic field even for high partial wave resonance. Therefore, the advantage for a high partial wave shape resonance is that a thorough investigation of strongly interacting quantum gas there becomes much more experimentally accessible.

Here we first report that a $d$-wave resonance is identified in ${}^{41}$K gas through atom loss. Very recently a similar $d$-wave Feshbach resonance has also been observed in ${}^{85}$Rb-${}^{87}$Rb mixture through atom loss as well\cite{YouLi}. In this work we present a systematic study of strongly interacting bosons near this $d$-wave resonance. We use magnetic field modulation spectroscopy to measure the molecular binding energy, and the results agree very well with multi-channel quantum defect theory (MQDT)~\cite{Gao1,Gao2,Gao3}. A few pieces of evidence are presented to show that this $d$-wave resonance is indeed a shape resonance. The lifetime of this strongly interacting bosons near the $d$-wave resonance is measured and found to be several tens of milliseconds, which is reasonably long. Many-body interaction effect has also been observed by ramping this Bose condensate through the resonance, during which a collective oscillation is excited. This is attributed to the change of the internal energy due to atom-molecule conversion, and a simple model is proposed to estimate the molecule fraction produced through the ramping.

\textbf {Observation of $d$-wave Resonance.} The experimental setup for preparing ultracold $^{41}$K gas has been described in our previous works~\cite{XingCanYao2016,Pan2017,Chen2016PRA}. Up to $7\times10^5$ ${}^{41}$K atoms in the $|F=1,m_F=1\rangle$ hyperfine state can be cooled to form a pure condensate at magnetic field $304.5$~G. The $^{41}$K Bose-Einstein condensate (BEC) is confined in a disk-like trap with radial and axial trapping frequencies being $\omega_\text{r}=20.8$~Hz and $\omega_\text{a}=84$~Hz, respectively. The Thomas-Fermi radii are about $28.8$~$\mu$m radially and $8$~$\mu$m axially.

To detect the scattering resonance, inelastic loss measurement is performed where enhanced atom losses occur at resonance due to a larger three-body decay rate. First, the magnetic field is rapidly ramped to $8$~G in $10$~ms and followed by a $100$~ms equilibrium time. Then, the magnetic field is ramped to the targeted value, and held there for a waiting time. Finally, to reduce the uncertainty in measuring atom number, we switch off the optical trap and ramp the magnetic field back to $304.5$~G simultaneously. After $20$~ms time-of-flight (TOF), strong saturation absorption imaging~\cite{Reinaudi2007OL} is performed to determine the remaining final number $N_\text{f}$. The accuracy of the magnetic field is calibrated by radio-frequency spectroscopy between the two lowest hyperfine states of $^{41}$K and it is found to be better than $10$~mG.

\begin{figure}[t]
  \centering
  \includegraphics[width=\columnwidth]{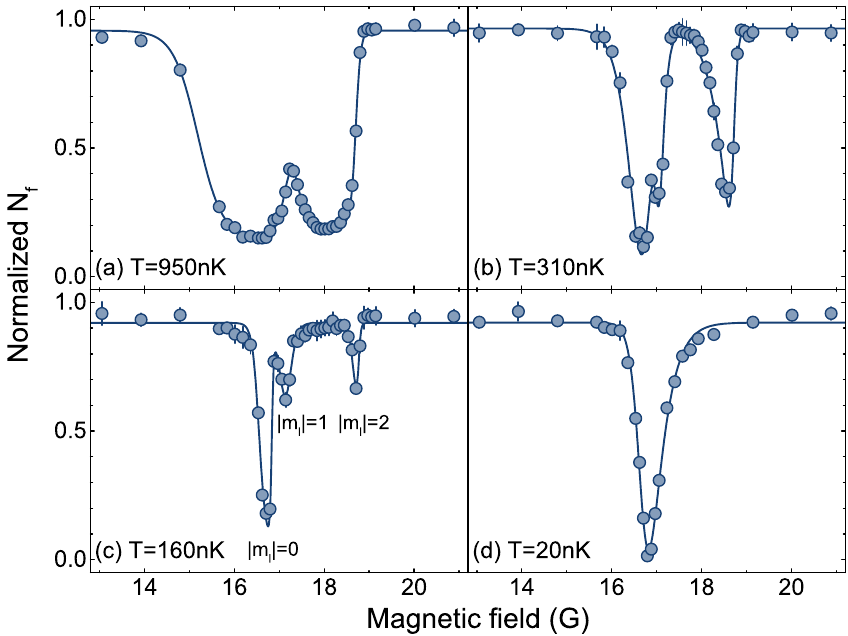}
  \caption{Inelastic loss spectroscopy of $^{41}$K atoms in the vicinity of d-wave shape resonance. (a)-(d) show the normalized remaining final number $N_\text{f}$ versus magnetic field at temperatures $950$~nK (a), $310$~nK (b), $160$~nK (c) and $20$~nK (d), respectively. Data points are counted and statistically averaged over four measurements with standard error margin being calculated (applicable to all figures in the following). The blue solid guiding lines in (a)-(d) are fits to the data points using asymmetric double Sigmoidal function.}\label{figure1}
\end{figure}

In Fig.~\ref{figure1} we show the remaining final number $N_\text{f}$ normalized by the initial atom number versus magnetic field for four different temperatures of $950$~nK, $310$~nK, $160$~nK, and $20$~nK, which respectively correspond to temperatures quite above BEC temperature ($T_\text{c}$), slightly below $T_\text{c}$, $0.85T_\text{c}$ and well below $T_\text{c}$ (i.e. a nearly pure BEC with no visible thermal fraction). In Fig.~\ref{figure1}(a), the loss curve shows two peaks and they display an asymmetric shape. This is one feature of the high partial wave resonance. As lowering the temperature below $T_\text{c}$, the thermal broadening becomes narrower, and one of the peak can now be resolved as two peaks, as shown in Fig.~\ref{figure1}(b) and (c). This is a hallmark of a $d$-wave resonance which can split into $m=0$, $m=\pm 1$ and $m=\pm 2$ resonances due to the magnetic dipolar interactions. It is easy to show that due to the dipolar energy splitting, the ratio of the distance between $m=\pm 1$ and $m=0$ resonances to the distance between $m=\pm 2$ and $m=0$ resonances is $4:1$, which is approximately consistent with the experimental data shown in Fig.~\ref{figure1}(b) and (c).

More intriguingly, as further lowering the temperature toward a pure BEC, two of the peaks associated with $m=\pm 1$ and $m=\pm 2$ disappear. This is because when the kinetic energy of atoms is too small to penetrate through the $d$-wave centrifugal barrier, the atomic loss induced by $d$-wave scattering is negligible. Nevertheless, because the dipolar interaction can couple the $s$-wave scattering state to a $d$-wave bound state, but only to the bound state with $m=0$ due to the rotational symmetry along $\hat{z}$, the resonance with $m=0$ can still lead to significant atom loss in very low temperature.

\begin{figure}[t]
\centering
\includegraphics[width=\columnwidth]{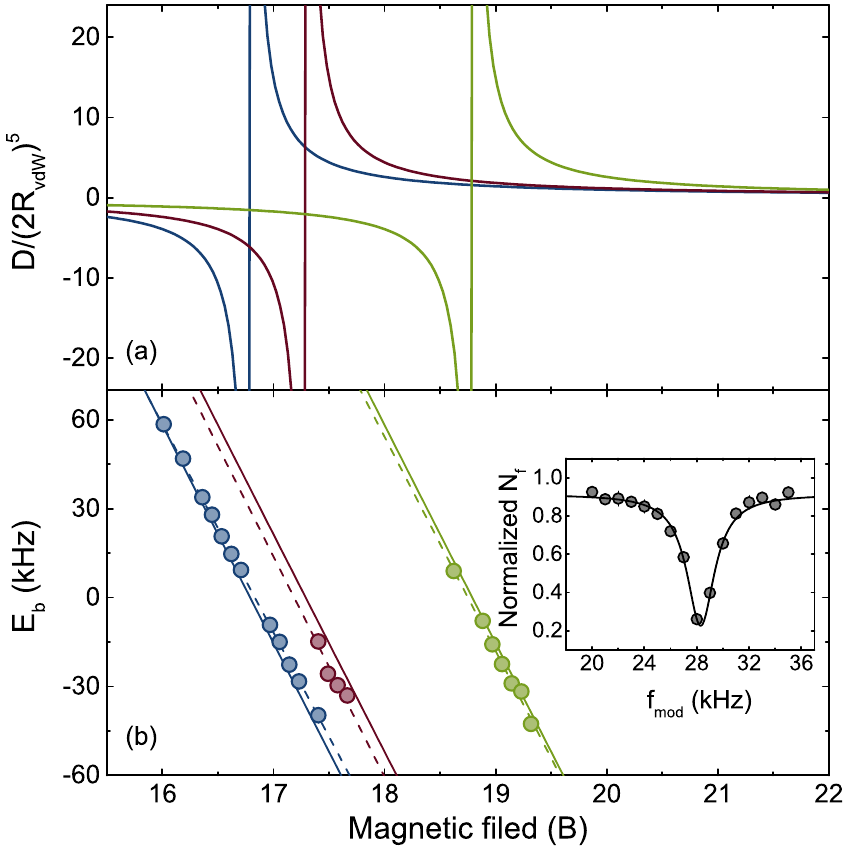}
\caption{Molecular binding energy of d-wave shape resonance. (a) The d-wave hyper-volume D normalized by $(2R_{\rm vdW})^5$ calculated by MQDT as a function of magnetic field, where $R_{\rm vdW}=65.42a_\text{0}$ is the van de Waals length of $^{41}$K atoms. (b) The measurement results (dots) and theoretical results from MQDT calculations (solid lines) of molecular binding energy as a function of magnetic field for $m=0$ (blue), $m=\pm 1$ (red) and $m=\pm 2$ (green) resonances, respectively. The dashed lines are the linear fittings of the experimental data. $E_b<0$ represents the bound state, while $E_b>0$ corresponds to the quasi-bound state. The inset shows a typical resonant association loss spectrum fitted with a Lorentz function.}
\label{figure2}
\end{figure}

\textbf{Measurement of Molecular Binding Energy.} Now we aim at a more direct detection of the molecules by applying a field modulation spectroscopy~\cite{Thompson2005PRL,Gaebler2007PRL}. After producing the condensate at $304.5$~G, we ramp the magnetic field to $15$~G in $10$~ms and hold at that field for an additional $100$~ms. Next, we ramp the magnetic field to a value near the resonance and then apply a small sinusoidal oscillation with frequency $f_\text{mod}$ for a duration between $200$~ms to $2$~s. To avoid amplitude dependent broadening and shifting effects, the modulation amplitude is chosen to be $50$~mG. For example, at magnetic field of $17.23$~G, as the modulation frequency $f$ varies, a resonant association loss spectrum of $^{41}$K BEC is obtained, yielding a resonant frequency of $28.30(3)$~kHz and spectrum width of $2.58(8)$~kHz, as shown in the inset of Fig.~\ref{figure2}.

Using this method, we can associate atoms not only into bound molecules but also into quasi-bound molecules. The binding energies $E_b$ are then extracted and plotted as a function of magnetic field, as shown in Fig.~\ref{figure2}. For $m=0$ resonance, a linear dependence of the binding energy on the magnetic-field is clearly observed with a slope of $B_0=70.36(58)$~kHz/G. The resonance position of $m=0$ is thus determined to be $16.833(3)$~G, which is in excellent coincidence with the peak loss position of $16.828(4)$~G obtained through inelastic loss spectrum of a pure BEC in Fig.~\ref{figure1}(d). The binding energy of molecular state associated with $m=\pm1$ is difficult to measure, mainly due to the small splitting between $m=0$ and $m=\pm1$ resonances. Therefore, we increase the modulation time to $5$~s and only observe association loss spectrum in the bound molecules side. The resonance position for $m=\pm1$ is measured to be $B_{\pm 1}=17.19(6)$~G, and the fitted slope is $74(12)$~kHz/G. The large error bars on these results are due to small data set of our measurement. For $m=\pm2$ resonance, the obtained resonance position is $B_{\pm 2}=18.75(1)$~G and slope is $72(2)$~kHz/G, respectively.

To compare experimental measurement with theory, we employ the MQDT calculation. We first exclude the dipolar interaction, and if we take the singlet and triplet scattering lengths $a_\text{S}=85.53a_\text{0}$ and $a_\text{T}=60.54a_\text{0}$ reported in previous literatures ($a_\text{0}$ is the Bohr radius) ~\cite{aSaT}, the MQDT predicts that the $d$-wave resonance locates at $B=65.1$~G. Here we fine tune the parameter $a_\text{T}$ to match the resonance position at $(B_0+B_{\pm 2})/2$, and we find $a_\text{T}=58.89a_\text{0}$. Using this new parameter of $a_\text{T}$, we then turn on dipolar interaction and compare the full results from MQDT with all experimental data sets. We can calculate both the $d$-wave super-volume $D$ and the bound state energy. The $s$-wave super-volume $D$ characterizes the $d$-wave scattering amplitude and a divergent $D$ locates the $d$-wave resonance. The results are shown in Fig.~\ref{figure2}, and we can see that the bound state energies predicted by MQDT agree well with the experiments.

\textbf{Evidence of Shape Resonance.} Here we report a few pieces of strong evidence because of which we are sure that the resonances observed here is actually a shape resonance.

(I) It is known that for a pure van de Waals potential at large distance $V(r)=-16R_{\rm vdW}^4/r^6$ ($m=\hbar=1$), a $d$-wave shape resonance occurs when the $s$-wave scattering length satisfies the condition \cite{Feshbach,Gao2}
\begin{equation}
a_s=0.956R_{\rm vdW}. \label{con}
\end{equation}
This condition is obtained by matching the solution of a van de Waals potential to the short range physics, the later of which is captured by the $s$-wave scattering length $a_\text{s}$. For two ${}^{41}$K atoms in their lowest hyperfine spin state $|F=1,m_F=1\rangle$, using the input parameters $a_\text{S}=85.53a_\text{0}$ and $a_\text{T}=58.89a_\text{0}$, our MQDT shows $a_\text{s}=62.30a_\text{0}$ at zero field. Knowing that $R_{\rm vdW}=65.42a_\text{0}$ for ${}^{41}$K, one has
\begin{equation}
a_s=0.952 R_{\rm vdW}. \label{reality}
\end{equation}
The already very small difference between Eq. \ref{reality} and Eq. \ref{con} tells that a $d$-wave bound state supported by this van der Waals potential is quite close to the scattering threshold. It is therefore conceivable that when the multi-channel coupling at short distance varies a little bit as tuning the magnetic field, it can cause a slow varying of $a_\text{s}$ and bring the system to the onset of a shape resonance.

(II) As shown in Fig.~\ref{figure2}, the bound state in our system appears at higher field and the quasi-bound state appears at lower field. This strong evidence demonstrates that the observed $d$-wave resonance is a shape resonance instead of a Feshbach resonance. That is because our atoms in the scattering states are in the lowest hyperfine spin state. If there were a Feshbach resonance, because atoms in the bound state come from a different hyperfine spin state, its energy must increase with respect to the energy of the scattering state, as the magnetic field increases. Thus, the bound state must sit at the lower field and the quasi-bound state must be at the higher field, which is in contrast to what we see here.

(III) From the MQDT calculation, one can also obtain the $\zeta_\text{res}$ parameter defined in \cite{Gao3}, where $|\zeta_\text{res}|\gg 1$ means the resonance is dominated by single channel. In our case, we find $\zeta_\text{res}=-201$ ($\zeta_\text{res}$ is always negative for all $\ell>0$ according to \cite{Gao3}). Our MQDT calculation also determines that the Bohr magneton for the bound state is $1.832$~MHz/G, very close to that of two atoms in the scattering state ($1.756$~MHz/G). Both shown that the bound state is dominated only by the scattering channel. In addition, close to a resonance (i.e., the one with $m=0$), we can cast $D$ as a function of $B$ into the form
\begin{equation}
D=D_{\text{bg}}\left(1+\frac{\Delta}{B-B_{0}}\right),
\end{equation}
where $D_\text{bg}$ is the background super-volume. Here we can formally define a resonance width $\Delta$ and it is found that $\Delta$ is as large as $180.3$~G. Although (III) can not distinguish a shape resonance from a broad Feshbach resonance, but as said, because a high partial wave Feshbach resonance is generically narrow due to the weak inter-channel coupling resulting from the centrifugal barrier, this can be taken as another evidence of shape resonance.

\begin{figure}[t]
  \centering
  \includegraphics[width=\columnwidth]{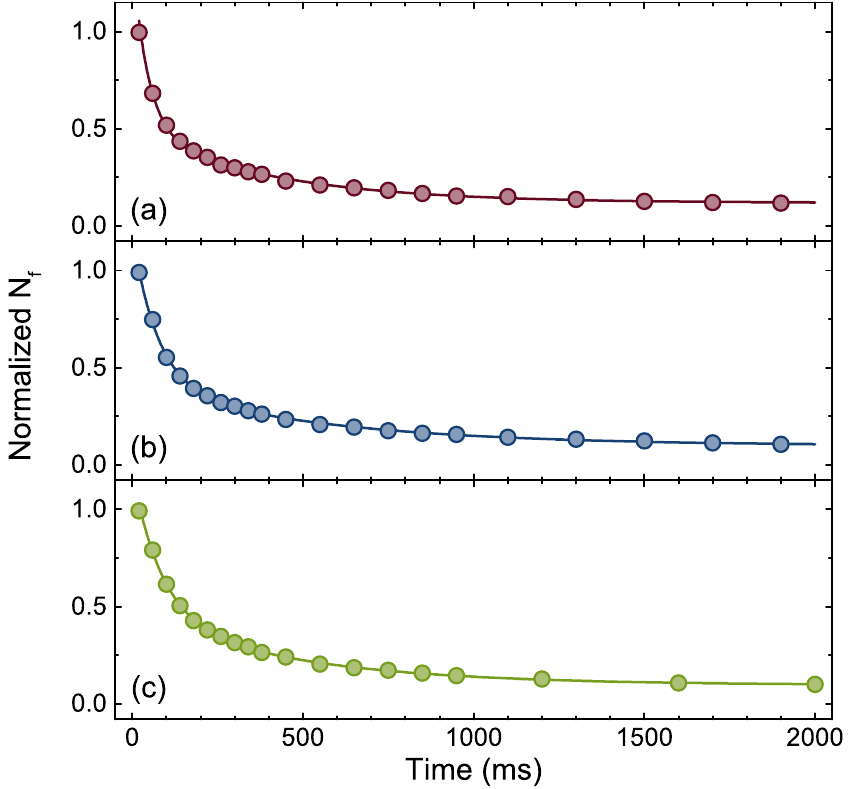}
  \caption{ Lifetime of $^{41}$K atoms in the vicinity of shape resonance. The final number normalized by the initial number is plotted as a function of holding time. The measurements are performed at magnetic field of $16.751$~G (a), $16.708$~G (b), and $16.621$~G (c), respectively. Solid curves correspond to fitting with a sum of two exponential functions. }\label{figure3}
\end{figure}

\textbf{Measurement of Lifetime.} Lifetime is very crucial for experimental study of this new strongly interacting quantum system. To measure the lifetime in the vicinity of this shape resonance, the cloud is cooled to about $300$~nK slightly below $T_c$, then we set the magnetic field to the quasi-bound side of resonance and measure the remaining final number $N_\text{f}$ as a function of holding time. We note that in the absorption detection, since the binding energy is negligible small, both atoms and possible molecules are counted in $N_\text{f}$.

$N_\text{f}$ versus holding time are plotted in Fig. \ref{figure3} for three different magnetic field values, where the binding energies of quasi-bound molecule roughly correspond to $k_\text{b}T_\text{c}$, $1.5k_\text{b}T_\text{c}$ and $2.5k_\text{b}T_\text{c}$, respectively. We find the curves in Fig. \ref{figure3} can be well fitted with a sum of two exponential functions as $\alpha_1 e^{- t/\tau_1}+\alpha_2 e^{-t/\tau_2}$, and the fitting results are $(\tau_1,\tau_2)=(44(2),401(14)) \text{ms}$ for Fig. \ref{figure3}(a), and $=(73(2),546(26)) \text{ms}$ for (b) and $=(84(3),464(24)) \text{ms}$ for (c). This lifetime is reasonably long, in particular, for the case of Fig. \ref{figure3}(a) sufficiently close to resonance. This provides an excellent opportunity to study many interesting strongly interacting physics in this system.

\begin{figure}[t]
  \centering
  \includegraphics[width=\columnwidth]{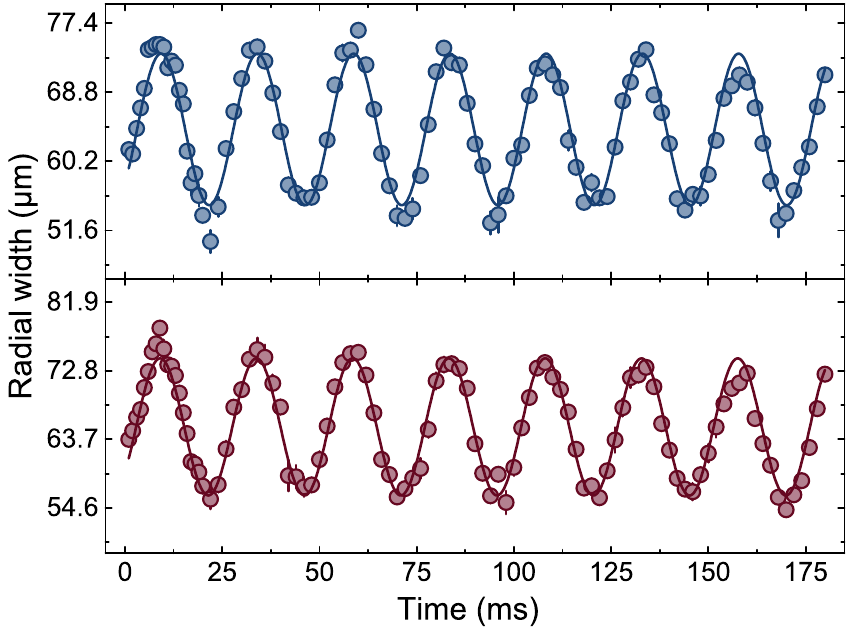}
  \caption{Collective oscillations of the radial width of the BEC after magnetic field ramping across the shape resonance. Blue (top) and red (bottom) circles correspond to the upward and downward magnetic field ramping across the resonance, respectively. Solid lines are the fitting curves based on a sinusoid model.}\label{figure4}
\end{figure}

\textbf{Collective Oscillation.} Finally we aim to observe interaction effect in this system, by starting with a pure BEC and performing a ramping through the $m=0$ resonance. We either start from an initial field $16.1$~G at quasi-bound side and upward ramp to the final field $17.056$~G at bound side with a speed $36.3~\text{G}/\text{s}$, or start from an initial field $17.404$~G at the bound side and downward ramp to the final field $16.621$~G at quasi-bound side with a speed $35.2~\text{G}/\text{s}$. The accuracy of the ramping velocity is guaranteed by measuring the real-time magnetic field with a fluxgate magnetometer. In the ramping process, there exists a certain period of time that the molecular binding energy is nearly degenerate with the energy of atoms, during which the atom-molecule conversion takes place. After the magnetic field ramping, the cloud is held at the final magnetic field for a variable time. Finally, the optical trap is switched off and the cloud is probed at $304.5$~G after $20$~ms TOF using the same imaging method described above. A striking feature is that a coherent oscillation of cloud radius are found in both cases, as shown in Fig.~\ref{figure4}. Both curves can be nicely approximated by a single frequency harmonic oscillations with small damping rates. The fitted oscillation frequencies $\omega_\text{c}$ are $40.56(10)$~Hz and $40.48(10)$~Hz for upward ramping and downward ramping, respectively. This gives $\omega_\text{c}=1.95\omega_\text{r}$, where $\omega_\text{r}$ is the radial trapping frequency introduced earlier. In our case, the trapping aspect radio $\lambda=\omega_\text{a}/\omega_{r}\approx 4$, the breathing mode frequency is approximately $1.81\omega_\text{r}$ \cite{breath1,breath2}. Thus we think this mode is consistent with a breathing mode.

\begin{figure}[t]
\centering
\includegraphics[width=\columnwidth]{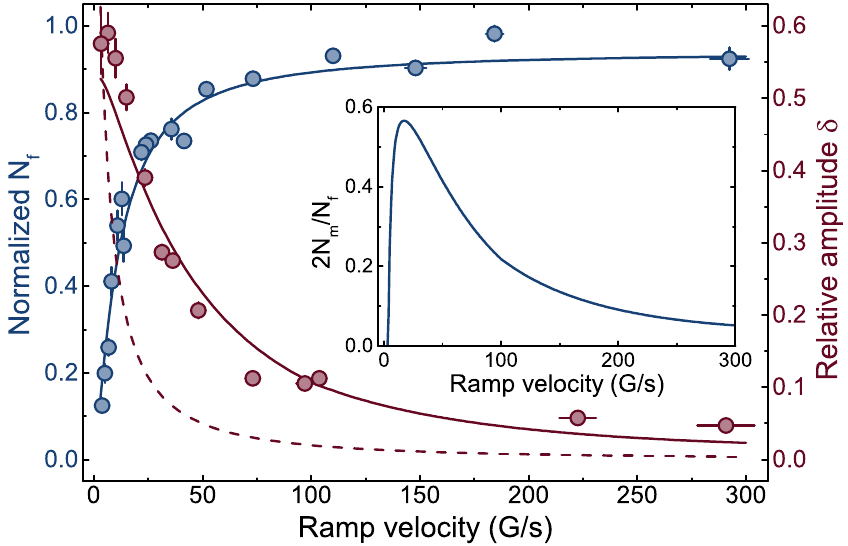}
\caption{The normalized remaining final  number $N_f$ (blue circles) and the relative oscillation amplitude $\delta$ (red circles) as a function of ramping velocity $v$. Blue and red solid lines are the fitting curves of $N_f$ and $\delta$ respectively. The dashed line shows the estimated $\delta$ assuming that the final cloud is purely made of atoms (see text for details). The discrepancy between the dashed line and the experimental data strongly supports the final cloud is a mixture of atoms and molecules. 
The inset shows the molecule fraction $2N_\text{m}/N_\text{f}$ as a function of ramping velocity $v$ extracted from a simple theoretical model.}
\label{figure5}
\end{figure}

A natural scenario attributes the excitation of the collective oscillation after ramping to the change of internal energy due to the atom-molecule conversion. To investigate this effect more clearly, we further take the case of downward ramping from $17.404$~G to $16.621$~G, and vary the ramping velocity from $3~\text{G}/\text{s}$ to $300~\text{G}/\text{s}$. For each ramping velocity, the remaining final number $N_\text{f}$ is measured immediately at the end of ramp and the BEC fraction is found to be larger than $90\%$. Fig. \ref{figure5} shows both normalized $N_\text{f}$ and oscillation amplitude $\delta$ as a function of the ramping velocity. We find that normalized $N_\text{f}$ increases and the oscillation amplitude decreases as the ramping velocity increases.

In general, the remaining final number $N_\text{f}$ should contain both atoms in the scattering state (with their total number denoted by $N_\text{a}$) and molecules in the bound state (with their total number denoted by $N_\text{m}$), and $N_\text{f}=N_\text{a}+2N_\text{m}$. If we assume that $N_\text{f}$ are all contributed by atoms in the scattering state, i.e. $N_\text{f}=N_\text{a}$, we can determine the radial Thomas-Fermi radius of the final cloud as $R_{\rm f }=(15a_\text{s} a_{\text{ho}}^4N_{\text{a}})^{1/5}=(15a_\text{s} a_{\text{ho}}^4N_{\text{f}})^{1/5}$, where $a_{\text{ho}}=\lambda^{1/4}\sqrt{\hbar/m\omega_r}$. On the other hand, the initial Thomas-Fermi radius can be calculated from the total initial atom number $N_\text{i}$ as $R_{\rm i }=(15a_\text{s} a_{\text{ho}}^4N_{\text{i}})^{1/5}$. Since the breathing mode is excited because the initial cloud of radius $R_\text{i}$ is suddenly brought to a situation with new equilibrium radius $R_\text{f}$, the relative oscillation amplitude $\delta$, defined as the peak to the valley distance divided by the average, is naturally estimated by
\begin{equation}
\frac{\delta}{2}=\frac{R_{\rm i}-R_{\rm f}}{R_{\rm f}}=\left(\frac{N_\text{i}}{N_\text{a}}\right)^{1/5}-1. \label{delta_Na}
\end{equation}
If we take $N_\text{a}=N_\text{f}$ for different ramping velocities, we determine $\delta$ as a function of $v$ shown by the dashed line in Fig. \ref{figure5}. It is very clear that it systematically underestimates the oscillation amplitude. In other words, we overestimate the final Thomas-Fermi radius.

This in fact strongly suggests the final cloud is a mixture of atoms and molecules. The actual final cloud size should be smaller because part of $N_\text{f}$ is actually contributed by the bound molecules. Here we need to make the following two bold assumptions of this atom-molecule mixture: (i) The molecules always reside in the atom cloud so that final Thomas-Fermi radius after ramping $R_{\rm f}$ is still determined by the atoms. (ii) The effect of atom-molecule interaction on the Thomas-Fermi radius is insignificant and can be ignored. Thus, we assume that the relation $R_{\rm f }=(15a_\text{s} a_{\text{ho}}^4N_{\text{a}})^{1/5}$ still hold. Now we take a best fit of experimental data for oscillation amplitude versus ramping velocity to determine a function $\delta(v)$, as shown by the solid line in Fig. \ref{figure5}. Therefore, with this function and the experimental data of $N_\text{i}$, we can determine $N_\text{a}$ with Eq. \ref{delta_Na}, and the molecule number is then extracted as $N_\text{m}=(N_\text{f}-N_\text{a})/2$.

The molecule fraction in the final cloud $2N_\text{m}/N_\text{f}$ is plotted in the inset of Fig. \ref{figure5}. One can see that this molecule fraction is small for both small or large ramping velocity. This result is quite reasonable because at small ramping velocity, the molecules have enough time to decay into a deep bound state through three-body loss and to escape the trap; and for fast ramping velocity there is not enough time for atoms to penetrate through the centrifugal barrier and to covert into molecules. As a result, the molecule fraction reaches a maximum at some intermediate ramping velocity. Therefore, our analysis suggests that a ramping with moderate velocity produces a mixture of atoms and molecules, which should be in a condensate phase for the temperature as low as our system. 

\textbf{Outlook.} In summary, we have discovered a very broad $d$-wave shape resonance in degenerate ${}^{41}$K gas and studied the strongly interacting many-body system of bosons near that. First of all, we find from both atom loss and molecular binding energy measurements that the resonances split into three branches, which is the hallmark for $d$-wave molecules. Secondly, we find the lifetime is sufficiently long in the strongly interacting regime. Thirdly, by ramping through the resonance, we observe the many-body interaction effect, which also provides strong evidence that the system becomes a mixture of both atoms and molecules after ramping. In addition, the nearly undamped breathing mode oscillation strongly suggests the system remains at a low-temperature condensate phase. All the evidence point to that a $d$-wave molecular superfluid can exist in our system. Studying $d$-wave superfluid can shed light on similar systems such as $d$-wave superconductor found in many high-Tc and strongly correlated materials.  

\begin{acknowledgements}
We thank Cheng Chin and Bo Gao for very helpful discussions. This work has been supported by the NSFC of China, the CAS, and the National Fundamental Research Program (under Grant Nos. 2013CB922001, 2016YFA0301600, 2016YFA0301600).

\end{acknowledgements}

%

%

\end{document}